\documentstyle[preprint,eqsecnum,aps]{revtex}
\begin{document}
\draft
\title{\bf Less than 50\% sublattice polarization\\ in an insulating
$S=$3/2 kagom\'{e} antiferromagnet at T$\approx$0}
\author{S.-H.  Lee$^{1,3}$, C. Broholm$^{2,3}$, 
M. F. Collins$^4$, L. Heller$^4$, A. P. Ramirez$^5$, 
C. Kloc$^{5,6}$, E. Bucher$^{5,6}$, R. W. Erwin$^3$, and N. Lacevic$^{2}$}
\address{
$^1$University of Maryland, College Park, MD 20742 \\
$^2$Department of Physics and Astronomy, 
The Johns Hopkins University, Baltimore, MD 21218 \\
$^3$National Institute of Standard and Technology, Gaithersburg, MD 20899\\
$^4$Department of Physics and Astronomy, 
McMaster University, Hamilton, Ontario, L8S 4M1, Canada\\
$^5$Lucent Technologies, Bell Laboratories, Murray Hill, NJ 07974 \\
$^6$University of Konstanz, Konstanz 7750, Germany}
 
\maketitle
\begin{abstract}
We have found weak long range antiferromagnetic order
in the quasi-two-dimensional insulating oxide $\rm KCr_3(OD)_6(SO_4)_2$ 
which contains  Cr$^{3+}$ $S=$3/2 ions on a
kagom\'{e} lattice. In a sample with $\approx$76\% occupancy of
the chromium sites the ordered moment is 1.1(3)$\mu_B$ per chromium ion
which is  only 
one third of the N\'{e}el value $g\mu_BS=3\mu_B$. 
The magnetic unit cell equals the
chemical unit cell, a situation which is favored by inter-plane 
interactions. Gapless quantum spin-fluctuations ($\Delta/k_B <0.25$K)
with a bandwidth of 60K $>>$ T$_N$ = 1.6K are the dominant contribution
to the spin correlation function, ${\cal S}(Q,\omega )$ in the ordered phase.
\end{abstract}
\narrowtext

\section{Introduction}

More often than not, local moments in insulating magnets
acquire a finite expectation value, $<\vec{S}>$ at  low    
temperatures. The only known exceptions are certain quasi-one-dimensional
antiferromagnets (AFM) in which  coupling between one dimensional units
cannot overcome the tendency of each unit to form a
cooperative quantum mechanical singlet.  
It is of great  interest then  to identify and study higher dimensional 
``moment free'' insulating spin systems.
We have made some progress towards this objective by identifying
a quasi-two-dimensional magnet in which $|<\vec{S}>|$ is
only $0.4(1)\times S$, where $S$ is the spin quantum number of the 
magnetic ion.

The material in question is $\rm KCr_3(OD)_6(SO_4)_2$, a 
quasi-two-dimensional AFM in which $S=$3/2 Cr$^{3+}$ ions 
occupy weakly interacting kagom\'{e} lattices. Quasi-two-dimensional
AFM's containing spins on kagom\'{e} lattices are likely to 
have reduced values of $|<\vec{S}>|$ as a consequence of the weak connectivity 
and triangular motif of the lattice. For Heisenberg
spins for example  it is known that
$|<\vec{S}>|/S \rightarrow 1$ for $S\rightarrow \infty$ 
and $T\rightarrow 0$\cite{huse,huber,ritchey,shender,chandra,sac92,chu92,har92} whereas
$|<\vec{S}>|= 0$ for $S=1/2$\cite{singhuse}.
The critical value of S below which a moment free magnet exists 
at T=0 is however unknown\cite{sac92}. 

Only a few AFM kagom\'{e} systems
have been studied so far.
SrCr$_{9p}$Ga$_{12-9p}$O$_{19}$ 
is a quasi-two-dimensional oxide  containing 
antiferromagnetically interacting $S=3/2$ Cr$^{3+}$ ions in 
kagom\'{e} bilayers\cite{obradors,ramirez,martinez,uemura,shlee}.
The material exhibits predominantly dynamic spin correlations 
although a spin-glass-like transition at $k_B T_g/J \approx 0.03$ causes
condensation of weak two-dimensional static correlations with 
an ordered moment corresponding to $|<\vec{S}>|=0.63(5)S$ and
a correlation length of order twice the spin-spin 
separation\cite{shlee,shl96,bro90}.  There are also recent reports of
short range magnetic order in $\rm D_3OFe_3(OH)_6(SO_4)_2$\cite{harrison}.
In contrast
the Fe-jarosites AFe$_3$(OH)$_6$(XO$_4$)$_2$, with A=K, Na, Tl and X=S or Cr 
which realize a $S=$5/2 kagom\'{e} AFM all have been found to 
develop long range order at low temperatures
with ordered moments $|<\vec{S}>|\approx 0.7S$ \cite{tow86,Collins}.

In this paper we report experiments characterizing static and dynamic
spin correlations in the $S=$3/2 jarosite kagom\'{e} lattice
KCr$_3$(OD)$_6$(SO$_4$)$_2$. 
Our principal result is that
long range order similar to that found in the $S=5/2$ jarosite systems does
develop but the spin polarization, $|<\vec{S}>|$,  is only 0.4(1) times
the N\'{e}el value, $S$.
Quantum fluctuations account for the remaining spin of
the Cr$^{3+}$ ions. 
The bandwidth of the fluctuation spectrum is approximately 6 meV and we
place an upper bound of 0.03 meV on any gap in the spectrum.

\section{Experimental Details}
The neutron scattering measurements were carried out 
on thermal neutron triple axis spectrometers
at the NIST reactor. Measurements were taken with fixed incident
neutron energies of 13.7meV or 5meV using PG or Be as  filters in
the incident beam. Because important conclusions of this paper
come from absolute measurements of the neutron scattering cross section, 
details of how we normalize scattering intensities from a 
powder sample are given in appendix A.

A 2.8g powder sample of $\rm KCr_3(OH)_6(SO_4)_2$ and
a 6g powder sample  of KCr$_3$(OD)$_6$(SO$_4$)$_2$ were synthesized 
using previously published methods\cite {synt}. 
We denote these samples A and B respectively.
Powder neutron scattering confirmed that the samples were single phase 
with the hexagonal alunite structure and low temperature lattice parameters 
a=7.2371 \AA~ and  c=16.9544 \AA. Fig. \ref{diffpattern} shows $T=20$ K 
diffraction data from sample B with the Rietveld fit superimposed.
In sample A the $\rm Cr^{3+}$ occupancy was 100\% as determined from
the slope of the inverse high temperature Curie-Weiss susceptibility versus T. 
In sample B
prompt gamma activation analysis, refinement of powder neutron
scattering (see Table 1 and Fig. \ref{diffpattern}) and chemical analysis indicated $\rm Cr^{3+}$ occupancy
of 76(5)\% whereas high temperature susceptibility measurement indicated 90\%
Cr occupancy. Unfortunately problems with stoichiometry are not uncommon
in this complicated class of materials\cite{rip86}. All neutron
scattering experiments reported here were performed on the deuterated
sample B.

The chromium lattice in $\rm KCr_3(OD)_6(SO_4)_2$
is illustrated in Fig~\ref{figspin}.
A formula unit contains three kagom\'{e} planes of chromium ions stacked
along the c-axis.
Neighboring kagom\'{e} planes (one shown with
solid lines, another with dashed lines)
are displaced by $(\frac{2}{3},\frac{1}{3},\frac{1}{3})$. 
A, B and C indicate three sublattices of the
magnetic structure to be discussed later.
$\rm Cr^{3+}$ ions are coordinated
by six oxygen atoms  located at the vertices of trigonally 
distorted octahedra\cite{tow86,alunite}.
Neighboring octahedra share one oxygen atom which is expected 
to mediate the strongest exchange interaction in the system between
neighboring chromium ions separated by $d=3.624$ \AA~ in a kagom\'{e} plane.
Susceptibility measurements on samples A and B showed
Curie-Weiss temperatures, $\Theta_{CW}$,
of -70(5)K and -54(2)K respectively. If nearest neighbor exchange in the
kagom\'{e} lattices indeed dominate, this
corresponds to a bond strength 
$J=(3k_B|\Theta_{CW}|/\tilde{z}S(S+1))=1.2(1)meV$ where 
$\tilde{z}=4n_{Cr}/9$ is the average nearest neighbor coordination ($n_{Cr}$
is the number of chromium atoms per unit cell
determined from Rietveld analysis and listed in
Table 1).
Other potentially significant exchange interactions in order of 
the strengths expected are the interplane coupling (Cr-K-Cr, d=6.025(1)\AA ) 
and the second nearest intraplane interaction 
(Cr-O-K-O-Cr, d=6.268(1)\AA ).

\section{Experimental Results}

\subsection{Antiferromagnetic long-range-order}

Fig. \ref{cvandm} shows the temperature dependence of
magnetic neutron diffraction at $Q=(012)$ and specific heat, C/T, 
which reveal a magnetic phase transition at $T_N=1.8$ and $1.55$K
$\ll \Theta_{CW}$ for samples A and B respectively. 
The entropy change between 0.2K and 2.9K is only 
0.15$R$ln(2S+1) and 0.16$R$ln(2S+1)  respectively indicating that the 
ordered moment in these samples may be small and/or
there may be substantial short range order above the N\'{e}el temperature.
Note that in contrast to $\rm D_3OFe_3(OH)_6(SO_4)_2$\cite{harrison}
where impurities induce long range magnetic order, it is our cleanest sample, A,
which has the 
most pronounced specific heat anomaly at $T_N$. This indicates that
long range order is the clean limit low T state of  $\rm KCr_3(OD)_6(SO_4)_2$.

Fig. \ref{cstescans} shows low temperature magnetic diffraction 
data for sample B.
The data was taken without final energy analysis and with an 
incident neutron energy, $E_i=13.7$ meV, which far exceeds the energy range with appreciable inelastic 
magnetic scattering. The data 
therefore probes the equal time spin correlation function
${\cal S}(Q)=\int^\infty_{-\infty} {\cal S}(Q,\omega)\hbar d\omega$. The two frames 
show low temperature (T=0.4K) data   with two different
backgrounds subtracted. Using diffraction data from immediately above the
N\'{e}el temperature (T=3.6K) as a background yields simply resolution limited 
magnetic Bragg diffraction indicating that the low temperature 
phase has AFM order. 
From comparing the line shapes
of magnetic  and nuclear Bragg peaks we conclude that the AFM
correlation length exceeds 500 \AA. Magnetic Bragg peaks
occur at the same values of $Q$ as nuclear Bragg peaks which
implies that the chemical unit cell described above is also 
a unit cell for the magnetic structure. This implies that 
all triangles of spins which are connected by in-plane
Bravais vectors of the crystal have equivalent spin orientations
which we denote by A, B, and C.
To determine the actual spin directions and the
relative orientation of spins in neighboring kagom\'{e} planes
we compared the measured Bragg intensities to those
calculated for specific model structures. Table 2 makes the comparison to
the spin configuration shown in Fig. \ref{figspin} (a). In this structure, spins
labeled by A, B, and C
were taken to lie within the basal plane and to be rotated 120$^o$
with respect to one another. Subsequent kagom\'{e} layers
are oriented so that each interplane triangle of chromium ions, 
such as those shaded in Fig. \ref{figspin}, contains
all three spin orientations : A, B, and C.
This structure provides the best
account of the measured intensities out of 72 high symmetry 
spin configurations which were examined. The ordered moment 
determined from this analysis was $|M|=1.1(3)\mu_B$ per occupied chromium site
which if $g\approx 2$ and $S=3/2$ implies that $|<\vec{S}>|=0.4(1)S$. Note that 
the error bar quoted here includes the uncertainty associated with
other choices of the interlayer registry which also provide
a reasonable account of the diffraction data.
 
\subsection{Equal-time short-range spin correlations}

Since the long range ordered moment is small, sum-rules imply that we should 
find static or dynamic  short range order which accounts
for the remaining spin on Cr$^{3+}$ ions. This aspect of the problem 
is examined in  Fig. \ref{cstescans} (b) which shows diffraction data
similar to those discussed above, though this time using  T=50K$\approx \Theta_{CW}$ 
data for background subtraction. At T=50K
we expect that magnetic scattering is almost $Q$-independent in the limited 
$Q$-range of the experiment. This is also
the case for the thermally activated inelastic phonon scattering which
is present in the high temperature data as well\cite{phoncon}. 
In  Fig. \ref{cstescans} (b) these terms are  over-subtracted
but the $Q$--dependence of the difference data should follow that of the 
low temperature equal time correlation function ${\cal S}(Q)$.
For the purpose of comparisons to theory we determined the
base-line for ${\cal S}(Q)$ by requiring that the data (including 
magnetic Bragg peaks)
satisfy the total moment sum rule
$3\times\int dQ Q^2 {\cal S}(Q)/\int dQ Q^2=n_{Cr}(g/2)^2S(S+1)$ using 
$n_{Cr}=6.9$ as derived from powder diffraction.
The most striking feature of the data is the  unusually small
fraction of the equal time correlation function which is associated with
magnetic Bragg peaks. The result indicates that 
most of the chromium spin 
in this material remain spatially disordered down to T=0.4K. 

\subsection{Energy resolved magnetic neutron scattering}

It is important to determine whether the short-range-order 
is static or dynamic. To
settle this question we used the full triple axis configuration
of the neutron spectrometer to measure the low temperature (T=70mK)
spectrum of magnetic fluctuations at two values of $Q$. The result is shown
in Fig. \ref{conq}.  Contributions to the detector count rate not associated
with inelastic scattering from the sample were measured and
subtracted as described in the figure caption. Integrating the 
data over $\hbar\omega$ we obtained the values for 
${\cal S}(Q)$ shown with solid symbols in Fig. \ref{cstescans} (b). 
Inelastic scattering with
0.4meV$<\hbar\omega<10$meV clearly accounts for
most of the equal time correlation function and hence we may conclude
that the short range order in $\rm KCr_3(OD)_6(SO_4)_2$ is mostly dynamic
in origin. From the energy spectra shown in Fig. \ref{conq} we
estimate a magnetic bandwidth of approximately 6meV. In addition we 
can place an upper limit of 0.4meV on any gap, $\Delta$,  in the magnetic
excitation spectrum. A stricter constraint on $\Delta$ is derived from 
the specific 
heat data of Fig. \ref{cvandm}
where the absence of activated behavior for $T>0.25$ K
indicates that $\Delta > 0.03$ meV.

Further information on magnetic fluctuations is provided by  the
constant-$\hbar\omega$ scans shown in Fig. \ref{conhw}. The data display 
an asymmetric peak
centered close to the strongest AFM
Bragg point, $\vec{\tau}=(101)$. The peak progressively broadens 
as energy transfer is increased.  The sharp leading edge 
and gradual decrease in intensity with increasing $Q$
is indicative of low dimensional magnetic correlations\cite{warr}. 
Based on {\em inelastic} constant-$Q$ and constant-$\hbar\omega$ scans we 
can also evaluate the a {\em lower bound} on the
total fluctuating moment
$3\times\int{\cal S}(Q,\hbar\omega) Q^2 dQ\hbar d\omega /\int Q^2 dQ=15$ per unit cell.
This number is an appreciable fraction of the total 
moment sum  $n_{Cr}(g/2)^2 S(S+1)
\approx 26$ per unit cell which indicates that the spin system 
remains predominantly dynamic at low temperatures in this material.

\section{Discussion}

\subsection{Magnetic structure}

Having presented the experimental results we now turn to comparison 
with relevant theory. The type of magnetic order 
in $\rm KCr_3(OD)_6(SO_4)_2$ is surprising.
Quantum as well as classical theories of the kagom\'{e} magnet 
predict that the so-called $\sqrt{3}\times\sqrt{3}$ in-plane 
structure [Fig. \ref{figspin} (b)]
is favored over the  ``q=0'' structure which we observe.  In addition another
kagom\'{e} related material $\rm SrCr_{9p}Ga_{12-9p}O_{19}$ shows 
short range $\sqrt{3}\times\sqrt{3}$ type order\cite{shl96,bro90}. It is therefore 
surprising to find q=0 type AFM order in 
$\rm KCr_3(OD)_6(SO_4)_2$.  We believe that 
interplane coupling plays an important role in stabilizing this type
of order in jarosite systems.
As is apparent from Fig. \ref{figspin} (a), q=0 kagom\'{e} AFM layers
can be stacked to take advantage of ferro- as well as antiferromagnetic
nearest neighbor interlayer coupling, but
$\sqrt{3}\times\sqrt{3}$ layers cannot. The reason is that the 
$(\frac{2}{3},\frac{1}{3})$ in-plane translation between neighboring 
kagom\'{e} planes 
is incompatible with the periodicity of the 
$\sqrt{3}\times\sqrt{3}$ magnetic structure and it 
is therefore impossible to arrange for all
interplane triangles of spins to contain all
three spin orientations A, B, and C as required to satisfy interlayer
coupling. Indeed,
all jarosite kagom\'{e} systems studied so far have q=0
type in-plane AFM order.  Interplane coupling also
controls another  aspect of the three dimensional magnetic
structure : the length of the primitive
magnetic unit cell in the c direction: 
Three identical but shifted
q=0 layers  can be stacked to
satisfy AFM interplane coupling. For ferromagnetic
interplane coupling, however, it is favorable to include 
time reversed copies of each of these three layers
in the stack which leads to a six layer sequence and a 
doubling of the unit cell in the c-direction. Such unit cell doubling
is actually observed for $\rm KFe_3(OH)_6(SO_4)_2$\cite{tow86}.

\subsection{Comparison to spin wave theory}

We have compared our inelastic data to specific models
for dynamic correlations in the kagom\'{e} AFM.
Several authors have considered spin waves in a long ranged ordered
classical  kagom\'{e} magnet and there
are a few unusual features to note. As a consequence of 
a local continuous degree of freedom (the weather vane mode) 
spin wave theories to lowest order in $1/S$ yield a zero energy
mode of excitations for all $\vec{Q}$. Furthermore it can be shown that
all classical long range ordered ground states lead to the same
dispersion relation for the finite energy modes.
We used the results of Harris et al.\cite{har92}
to calculate the inelastic
neutron scattering cross section associated with the finite energy modes
of the =0 structure\cite{shlthesis}. 
The calculated cross section was  convoluted with the experimental
energy resolution function and is shown as dashed lines in Fig. \ref{conhw}.
The comparison involves only a single adjustable parameter which is the overall
scale factor. The exchange constant was fixed to 1meV, the value extracted from
high temperature susceptibility measurements.
Simple lowest order two dimensional  spin wave theory clearly cannot 
account well for the data. Specifically the sharp 
features predicted by the theory are absent in the data. 
This is  actually quite unusual
for two dimensional magnets. The qualitative features of
spin dynamics in
$\rm La_2CuO_4$ are for example very well described by such theories\cite{gabe}.  
\subsection{Equal-time spin correlations}

Theoretical predictions are also available for the
equal time correlation function ${\cal S}(Q)$. We compare these to 
our data in Fig. \ref{cstescans}.
The dotted line is the result of a Monte Carlo study~\cite{rei93},
and the dashed line resulted from a high-T expansion~\cite{har92}.
Given that the experimental  data probe {\em low temperature} spin correlations
it is surprising that apart from the magnetic Bragg peaks
the high temperature expansion accounts quite well for the data.
Evidently the frustration inherent to the kagom\'{e} AFM 
prevents much evolution of equal
time correlations between high and low temperatures in 
$\rm KCr_3(OD)_6(SO_4)_2$.

\section{Summary and Conclusion}

In summary, we have shown that long range AFM
order exists below a second order transition at T$_N$=1.8K 
in the $S=$3/2 jarosite kagom\'{e} system 
$\rm KCr_3(OD)_6(SO_4)_2$. The ordered structure
preserves the chemical unit cell and we have argued
that this type of structure is favored by interplane coupling.
The ordered moment per Cr ion in a sample with 76\% occupancy on the
chromium sublattice 
is 1.1(3)$\mu_B$ per occupied chromium site which is only one third of that corresponding to 
N\'{e}el order for $S=$3/2 spins. Concomitantly spin fluctuations
are unusually strong in this system and cannot be accounted for
by conventional spin wave theory. The inset to Fig. \ref{conq} summarizes
$|<\vec{S}>|/S$ versus $S$ for various kagom\'{e} systems studies so far.
The data indicate that the weakly diluted $S=$3/2 kagom\'{e} AFM 
may be close to a T=0 quantum critical point separating 
a phase with $|<\vec{S}>|\ne 0$ from a phase with $|<\vec{S}>| = 0$ \cite{sac92}.

\acknowledgements
We acknowledge stimulating discussions 
with Drs. A. J. Berlinsky, C. Kallin, M. Townsend,
J. E. Dutrizac and A.Chubukov. Work at JHU was supported  
by the National Science Foundation through DMR-9302065 and 
DMR-9453362. CB is grateful for the hospitality of the ICTP
in Trieste where parts of this manuscript were prepared.

\appendix

\section{Normalizing Powder Neutron Scattering Data}

\subsection{Neutron Scattering Cross Sections}

The magnetic neutron scattering cross section from a powder sample 
at wave vector transfer, $Q$, and energy
transfer, $\hbar\omega$ can be written\cite{lovesey}
\begin{equation}
\frac{{\rm d}^2\sigma}{{\rm d}\Omega{\rm d}E^{\prime}}=
\frac{k^{\prime}}{k}N r_0^2 e^{-2W(Q)}
~2 {\cal S}(Q,\omega )
\end{equation}
where $r_0=5.38$ fm, 
$e^{-2W(Q)}$ is the Debye Waller factor (which we generally 
take to be unity at low $Q$ and low $T$),
and ${\cal S}(Q,\omega )$ is the
spherically averaged scattering function:
\begin{equation}
{\cal S}(Q,\omega)=\int \frac{d\Omega_{\hat{Q}}}{4\pi} \frac{1}{2}
\sum_{\alpha\beta} (\delta_{\alpha\beta}-\hat{Q}_\alpha\hat{Q}_\beta)
S^{\alpha\beta}(\vec{Q},\omega).
\end{equation}
The scattering function is given by
\begin{equation}
\label{eq_sqw}
{\cal S}^{\alpha\beta}(\vec{Q},\omega)=|\frac{g}{2}f(\vec{Q})|^2 \frac{1}{2\pi\hbar}
\int dte^{i\omega t}\frac{1}{N} \sum_{\vec{R}\vec{R}^{'}}
<S^\alpha_{\vec{R}}(t)S^\beta_{\vec{R}^{'}}(0)> 
e^{-i\vec{Q} \cdot (\vec{R}-\vec{R}^{'})}
\end{equation}
where $f(\vec{Q})$ is the magnetic form-factor of a single
Cr ion\cite{crff}. 

The  Bragg scattering cross section 
for a powder sample takes the form  
\begin{equation}
\frac{{\rm d}\sigma}{{\rm d}\Omega}=N\sum_{\tau}{\cal I}(\tau )\delta (Q-\tau )
\end{equation}
where the summation is over reciprocal lattice vectors, $\tau$, with different
lengths. The 
$Q$--integrated nuclear scattering intensity is
\begin{equation}
{\cal I}_N(\tau )=\frac{v^*}{4\pi\tau^2}\sum_{|\vec{\tau}^{~\prime}|=\tau}
|F_N(\vec{\tau}^{~\prime})|^2
\label{inuc}
\end{equation}
where $v^*$ is the volume of the reciprocal lattice unit cell
and the structure factor is
\begin{equation}
F_N(\vec{Q})=\sum_{\vec{d}}
e^{-W_{\vec{d}}(\vec{Q})}
b_{\vec{d}}~e^{i\vec{Q}\cdot\vec{d}}
\end{equation}

For magnetic Bragg scattering  the $Q$--integrated intensity is
\begin{equation}
{\cal I}_m(\tau )=r_0^2\frac{v^*}{4\pi\tau^2}
\sum_{|\vec{\tau}^{~\prime}|=\tau}
|\frac{g}{2}f(\vec{\tau}^{~\prime})|^2
(|\vec{F}_M(\vec{\tau}^{~\prime} )|^2-|\hat{\tau}\cdot\vec{F}_M(\vec{\tau}^{~\prime} )|^2),
\label{imag}
\end{equation}
where
the magnetic vector structure factor is given by
\begin{equation}
\vec{F}_M(\vec{Q} )=\sum_{\vec{d}}e^{-W_{\vec{d}}(\vec{Q})}
<\vec{S}_{\vec{d}}>e^{i\vec{Q}\cdot\vec{d}}
\end{equation}

\subsection{Normalizing the measured Intensities}

The detector count rate is related to the 
scattering cross section through an expression of the form:
\begin{equation}
I(Q,\hbar\omega )= {\cal C}(\hbar\omega )
\int dQ^{\prime} \hbar d\omega ^{\prime}
{\cal R}_{Q\omega }(Q-Q^{\prime},\omega-\omega^{\prime})
\frac{{\rm d}^2\sigma}{{\rm d}\Omega{\rm d}E^{\prime}}
(Q^{\prime},\hbar\omega^{\prime}),
\end{equation}
where the spectrometer resolution function is normalized to unity:
\begin{equation}
\int dQ^{\prime} \hbar d\omega^{\prime} {\cal R}_{Q\omega}(
Q-Q^{\prime},\omega-\omega^{\prime})=1
\end{equation}
The resolution function and the 
energy dependence of ${\cal C}(\hbar\omega)/{\cal C}(0)$ can be calculated 
given the spectrometer set up\cite{chesseraxe}. For experiments
with fixed incident energy it is a good approximation
to use
\begin{equation}
\frac{{\cal C}(\hbar\omega)}{{\cal C}(0)}=(\frac{k^{\prime}}{k})^2
\sqrt{\frac{(2k^{\prime})^2-\tau_A^2}{(2k)^2-\tau_A^2}},
\end{equation}
where $k$ and $k^{\prime}$ are the incident and scattered
neutron wave vectors, and $\vec{\tau}_A$ is the reciprocal lattice vector of 
the Bragg reflection
being used as analyzer.
The value 
${\cal C}(0)$ depends on many factors such as 
the neutron flux, monochromator and analyzer reflectivities,  collimator transmissions, and detector
efficiencies. We determined ${\cal C}(0)$
by measuring the $Q$--integrated intensity of nuclear Bragg peaks
with known values of ${\cal I}_N(\tau )$:
\begin{equation}
N{\cal C}(0)=\frac{\int I_{\rm Bragg}(Q-\tau,\hbar\omega ) dQ\hbar d\omega}{{\cal I}_N(\tau )}
\end{equation}
With this number it is straight forward to derive absolute values
of ${\cal I}_M(\tau )$ and ${\cal S}(Q,\hbar\omega )$ from the
detector count rates associated with magnetic scattering:
\begin{eqnarray}
{\cal I}_M(\tau ) &=& \frac{1}{N{\cal C}(0)} \int I_{\rm Bragg}(Q,\hbar\omega ) 
dQ\hbar d\omega \\
{\cal S}(Q,\omega ) &\approx& \frac{1}{N{\cal C}(0) r_0^2} (\frac{k}{k^{\prime}})^3
\sqrt{\frac{(2k)^2-\tau_A^2}{(2k^{\prime})^2-\tau_A^2}}~
\frac{1}{2}I(Q,\hbar\omega ).
\end{eqnarray}
Resolution smearing was neglected in the last expression.

\begin{table}
\caption{Positions within space group R${\bar 3}$m  
and occupancies per unit cell
of atoms in $\rm KCr_3(OD)_6(SO_4)_2$
at T=20K. The numbers were derived from the
data shown in Fig. 1 by Rietveld analysis 
using GSAS\protect\cite{gsas}. 
Isotropic Debye-Waller factors, $\exp (-<u^2>Q^2 )$,
were used and the resulting overall reduced $\chi^2=3.8$.}
  \begin{tabular}{lllllll} 
&	Site&	x	&y	&z	&n/f.u. & $\sqrt{<u^2>}/\AA$\\ \hline
K&	3b	&0	&0	&0	&3 & 0.090(5) \\
Cr&	9d	&0	&0.5	&0.5	&6.9(2) & 0.072(6)\\
O&	18h	&0.1282(2)&0.2564(3)&0.1341(1)	&16.8(3) & 0.072(2)\\
D&	18h	&0.1961(2)&0.3922(5)&0.1084(2)	&13.8(3) & 0.106(2)\\
H&	18h	&0.1961(2)&0.3922(5) &0.1084(2)&4.2(1)& 0.106(2) \\
S&	6c	&0 &0	&0.3061(4)	&5.73(9) & 0.065(5) \\
O&	6c	&0 & 0 &0.3903(2)&5.8(2)&0.075(3) \\
O&	18h	&0.2218(2)&0.4436(3)&$-$0.0573(1)&17.4(3) &0.077(2) 
  \end{tabular}
\end{table}

\begin{table}[htb]
\caption{The measured and calculated $Q$--integrated intensities of
Bragg peaks for $\rm KCr_3(OD)_6(SO_4)_2$ powder
sample B. Magnetic and nuclear peaks were measured with the same 
experimental configuration. 
Magnetic Bragg intensities  were
obtained from the difference between the scattering intensity at 0.4K
and 3.6K  shown in Fig. \protect\ref{cstescans} (a).   
${\cal I}_M^{\rm cal}(\tau)$ was calculated using 
Eq. (\protect\ref{imag}) for the
spin configuration shown in Fig. \protect\ref{figspin} with
$(g/2)<\vec{S}_{\vec{d}}>=0.55$. Comparing measurement to calculation
we obtain a residual coefficient 
$R=\sum_\tau |{\cal I}^{obs}(\tau)-{\cal I}^{cal}(\tau)|/
\sum_\tau{\cal I}^{obs}(\tau)=0.17$. For nuclear Bragg peaks we got
$R=0.14$ when comparing  ${\cal I}_N^{\rm obs}(\tau)$
to ${\cal I}_N^{\rm cal}(\tau)$. The latter values were calculated from
Eq. (\protect \ref{inuc}) and Table 1.}

\label{magbrag}
\begin{tabular}{cccccc}  
$\vec{\tau}$&$Q^{\rm obs}$& ${\cal I}_M^{\rm obs}(\tau)$  & 
${\cal I}_M^{\rm cal}(\tau)$ 
& ${\cal I}_N^{\rm obs}(\tau)$ & ${\cal I}_N^{\rm cal}(\tau)$\\
(h,k,l) & (\AA$^{-1}$)& ($10^{-18}$ m) & 
 ($10^{-18}$ m) 
&($10^{-18}$ m) &($10^{-18}$ m)\\
\hline
  1  0  1 & 1.070 &    0.061(6) &   0.0581  &   0.32(3) &   0.313\\
  0  0  3 & 1.115 &      -      &      -    &   2.21(4) &    1.029\\
  0  1  2 & 1.240 &    0.038(6) &   0.0419  &   0.03(3) &    0.203\\
  1 1 0 & 1.731 &      -      &   0.0065  &  0.06(1)  &   0.165\\
  1  0  4 & 1.791 &    0.019(6) &  0.0184   &  0.32(3)  & 0.261\\
   2 0 $\bar{1}$, 1  1  3 & 2.063 &   0.01(1)  &   0.0119  &   4.8(1)  &    4.683\\
  1 0 $\bar{5}$, 2 0 2 &2.139 &      -      &            &  4.2(1)  &   3.632\\
  0 0 6 &2.222&      -      &     -     &  0.17(5)  &  0.158\\
  2 0 $\bar{4}$ &2.498&      -      &     -     &   0.42(1) &    0.194\\
  2 0 5, 2 1 $\bar{2}$,1 0 7 &2.753&      -      &      -    &   14.45(6)&     14.057\\
  3 0 0, 2 1 4 &3.018&      -      &     -     &  1.16(3)  &   0.977\\
  3 0 $\bar{3}$, 2 1 $\bar{5}$ &3.216&      -      &      -    &  2.58(6)  &    3.667\\
  2 0 $\bar{7}$ &3.285&      -      &     -     &  1.45(6)  &    1.710\\
  2 2 0 &3.476&      -      &      -    &   2.9(1)  &    3.290\\

\end{tabular}

\end{table}

\begin{figure}
\caption{Powder diffraction data from sample B at $T=20$ K
taken on the 32 detector 
powder diffractometer BT 1 at NIST. We used the Cu (311)
monochromator with $\lambda=1.540$ \AA~ and collimations $15^\prime-20^\prime-7^\prime$. The line through the data shows
the Rietveld fit which led to the structural parameters shown in 
Table 1. The bottom panel shows the difference between model and data.}
\label{diffpattern}
\end{figure}

\begin{figure}
\caption{Stacked kagom\'{e} lattices as they occur in
jarosites. A, B, and C denote three magnetic sublattices.
(a) is the spin configuration we propose for $\rm KCr_3(OD)_6(SO_4)_2$
which satisfies inter-plane exchange interactions (shaded triangle).
(b) Shows that a stacking of kagom\'{e} layers with the  
$\protect\sqrt{3} \times\protect\sqrt{3}$ AFM structure 
cannot satisfy inter-plane coupling.
}
\label{figspin}
\end{figure}

\begin{figure}
\caption{(a) Temperature dependence of the magnetic Bragg
intensity at $Q=(012)$ which is proportional to the squared staggered
magnetization.
(b) Temperature dependence of the specific heat divided by
temperature for two samples with different Cr concentration in the 
kagom\'{e} layers.
A 100\% Cr$^{3+}$ and B: 75-90\% Cr$^{3+}$.}
\label{cvandm}
\end{figure} 

\begin{figure}
\caption{
Difference between low and high temperature neutron diffraction data.
The data were obtained with $E_i=13.7$meV, collimations 
$40^{'}-36^{'}-36^{'}$ and no analyzer.
(a) Shows I(0.4K)-I(3.6K) probing
spin correlations which develop between T=3.6K and T=0.4K. 
(b) Shows I(0.4K)-I(50K) which is a measure of 
the entire low temperature equal time spin correlation function.
The solid symbols were obtained by integrating the constant-$Q$ scans
shown in Fig. \protect\ref{conq}. 
Dotted and dashed lines are described in the text.}
\label{cstescans}
\end{figure}

\begin{figure}
\caption{
Constant-$Q$ scans at $Q=1.1$\AA$^{-1}$ (filled symbols) and
$Q=0.55$\AA$^{-1}$ (open symbols).  Triangles were obtained with
$E_i$=5meV and collimations $60^{'}-40^{'}-80^{'}-80^{'}$, circles with
$E_i$=13.7meV and $60^{'}-20^{'}-40^{'}-40^{'}$ and rectangles with
$E_i$=14.7meV and collimations $60^{'}-20^{'}-40^{'}-40^{'}$.
Backgrounds were subtracted as follows :  For $\hbar\omega \le 0.7$meV
we subtracted neutron energy gain data.  For $\hbar\omega > 0.7$meV we
subtracted data obtained with the analyzer rotated 10$^o$ from its
reflection condition.  The inset shows $|<\vec{S}>|/S$ versus S for
various kagom\'{e} systems.  Squares: $S= 5/2$ Fe-jarosite systems with
long range order \protect\cite{tow86}; Open circles: frozen moment in
the spin glass phase of $\rm SrCr_{9p}Ga_{12-9p}O_{19}$ with p=0.92 and
$S=3/2$ \protect\cite{shl96}; Filled circle:  long
range ordered moment in the $S=3/2$ Cr-jarosite
$\rm KCr_3(OD)_6(SO_4)_2$ with 76\% occupancy of the 
kagom\'{e} lattice.}

\label{conq}
\end{figure}

\begin{figure}
\caption{Constant-$\hbar\omega$ scans at 0.5, 1.1, and 2 meV.
The $\hbar\omega$=0.5meV scan was obtained with $E_i$=5meV and 
collimations $60^{'}-40^{'}-80^{'}-80^{'}$. 
The $\hbar\omega$=1.1meV and $\hbar\omega$=2.0meV
scans were obtained with $E_i$=13.7meV and collimations
$60^{'}-20^{'}-40^{'}-40^{'}$. The background subtraction 
scheme is described in the caption to Fig. \protect\ref{conq}. 
Solid lines are a single parameter
fit to a spin wave theory described as in the text.}
\label{conhw}
\end{figure}


\begin{references}
\bibitem{phoncon}The one phonon scattering cross section rises as $Q^2$.
However, for a powder sample in the small $Q$ limit 
inelastic phonon scattering mainly
occurs in combination with an incoherent elastic 
scattering event resulting in an almost $Q$-independent 
double scattering process.
\end{references}
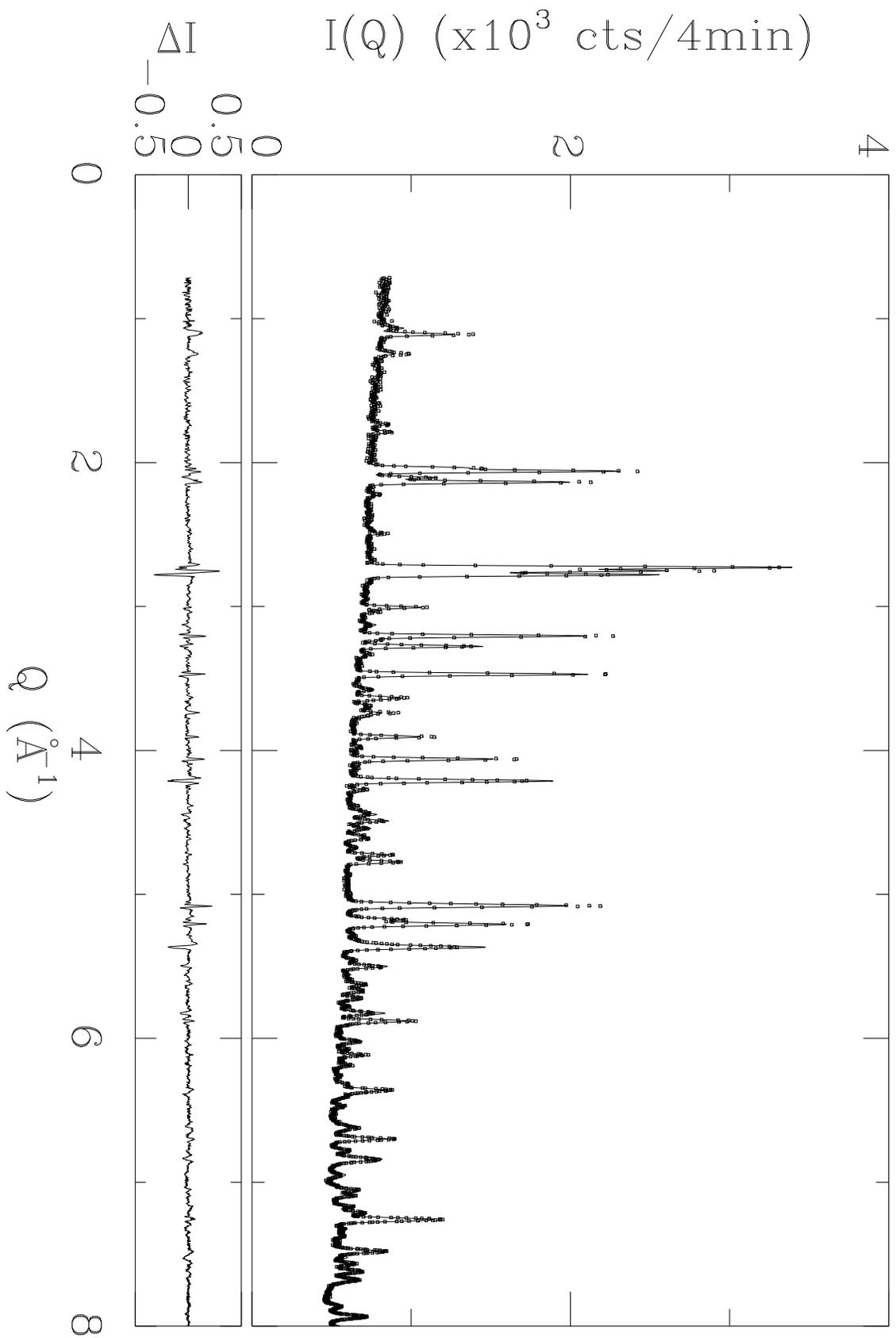
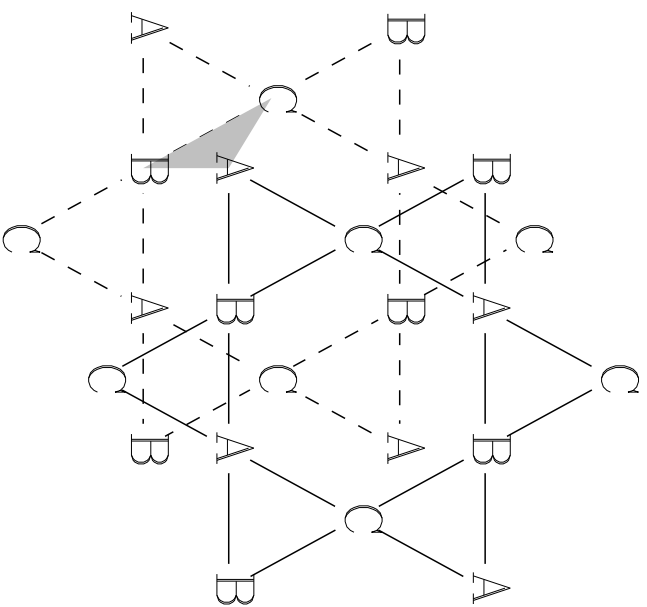
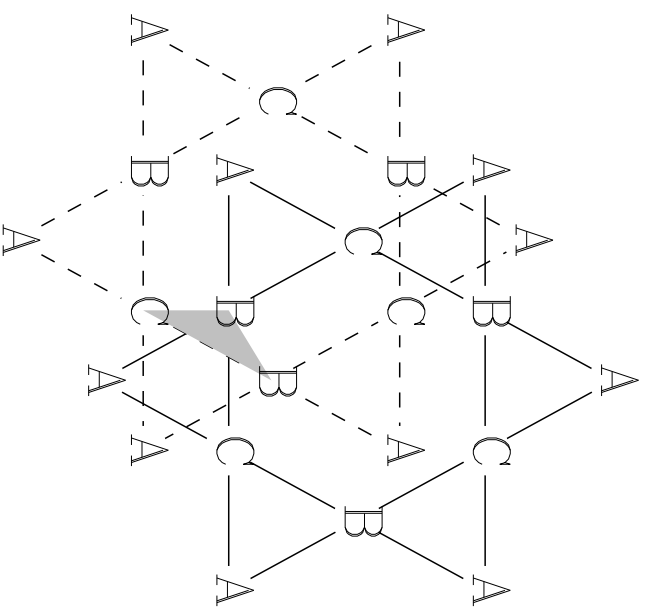
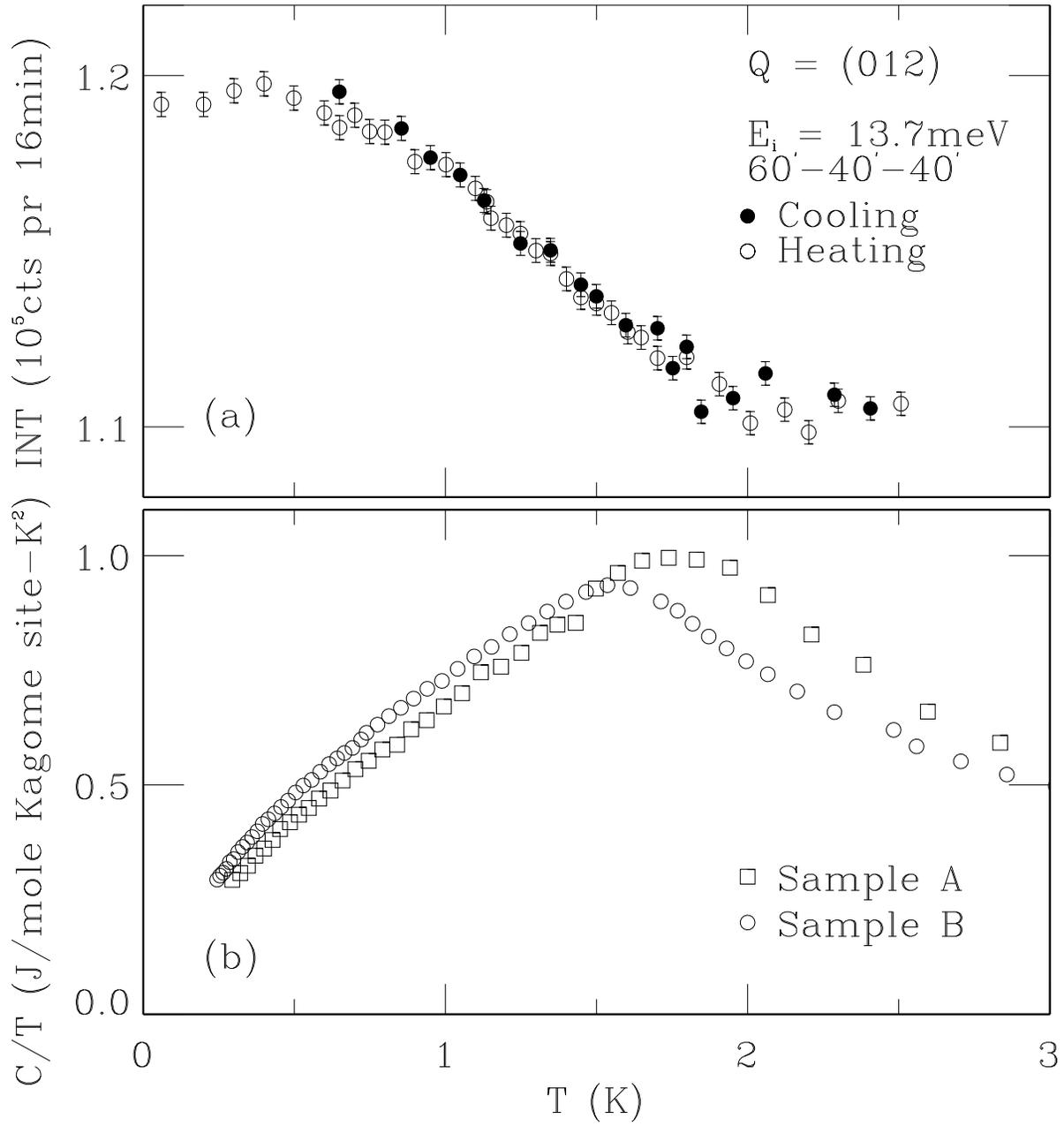
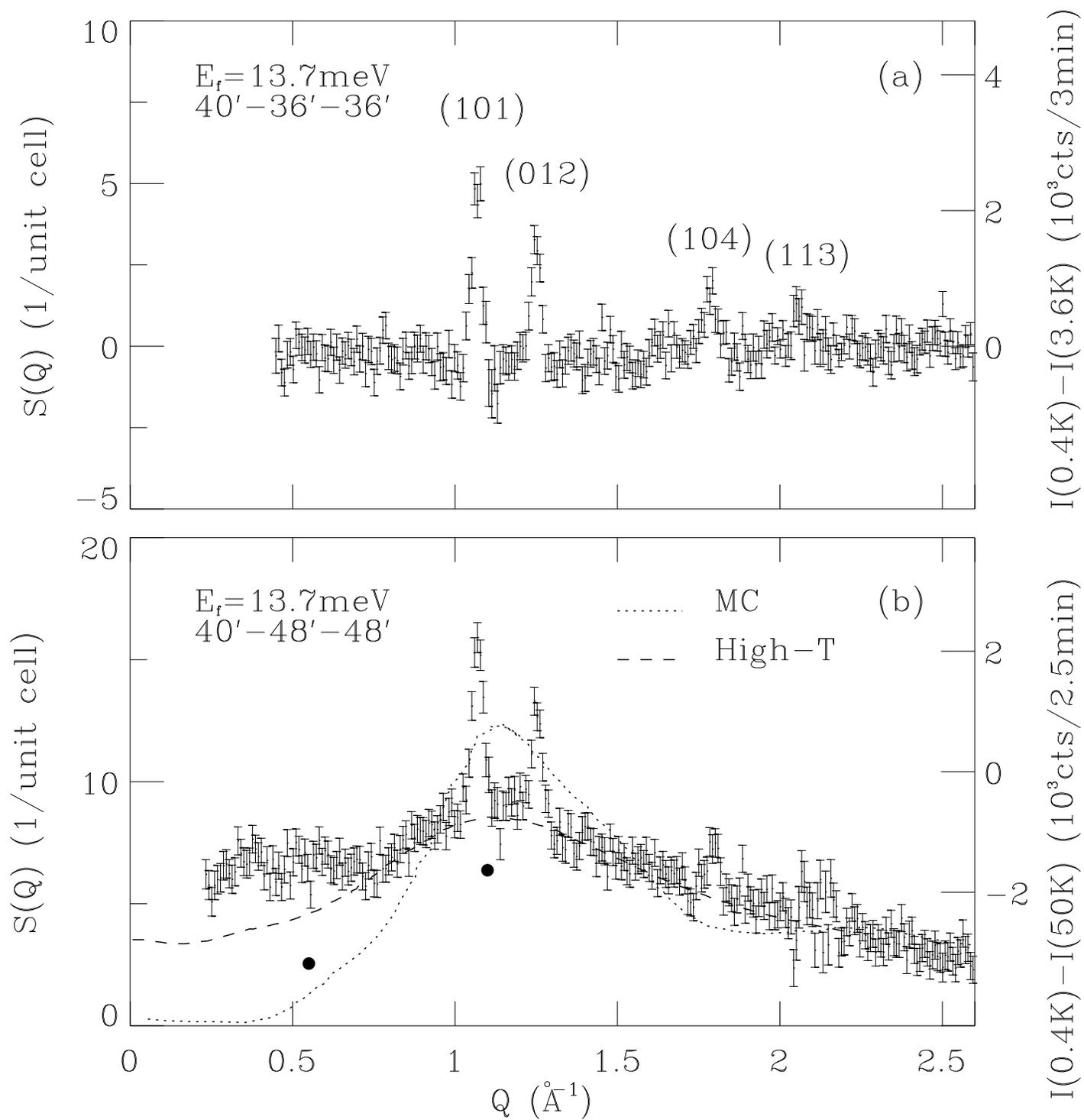
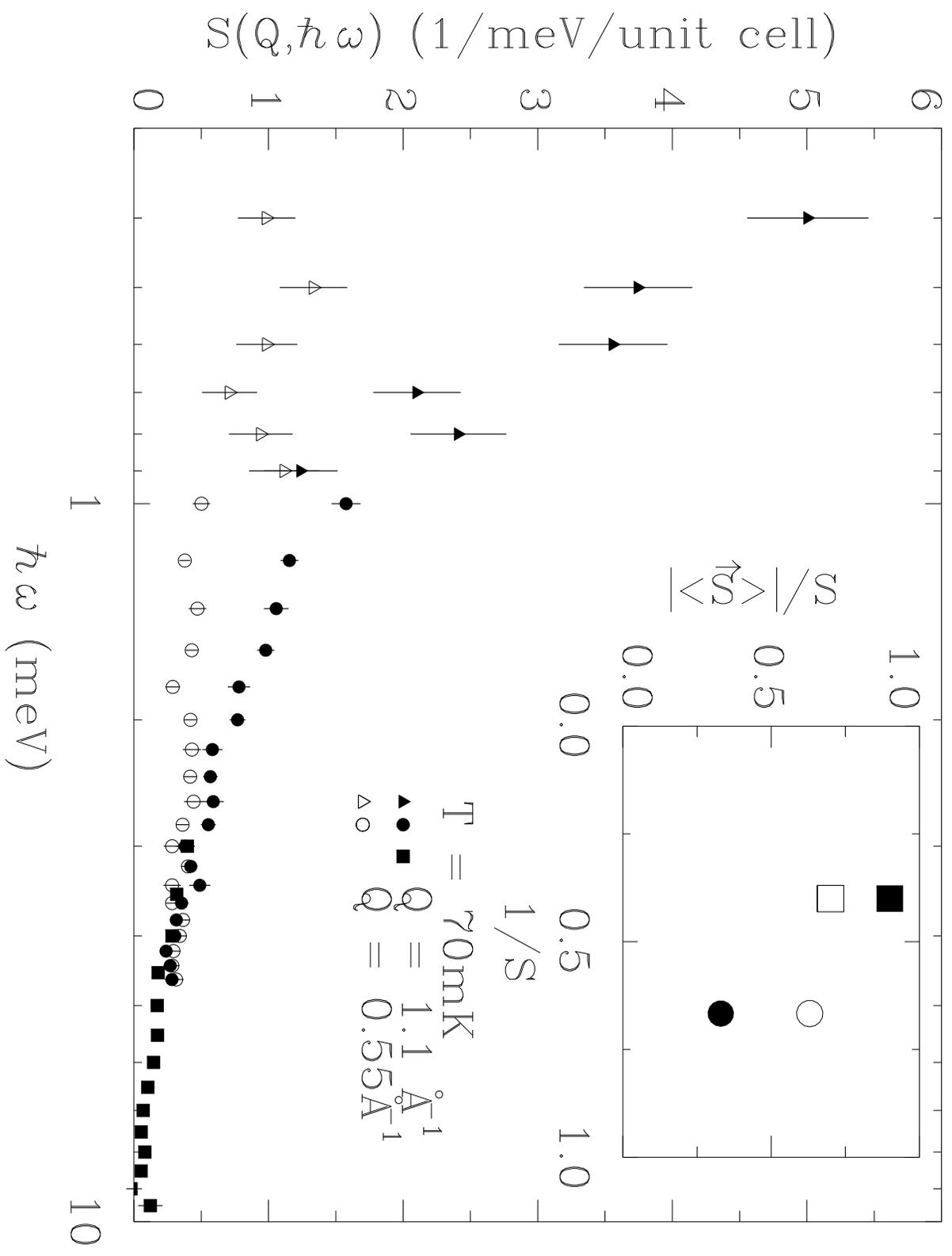
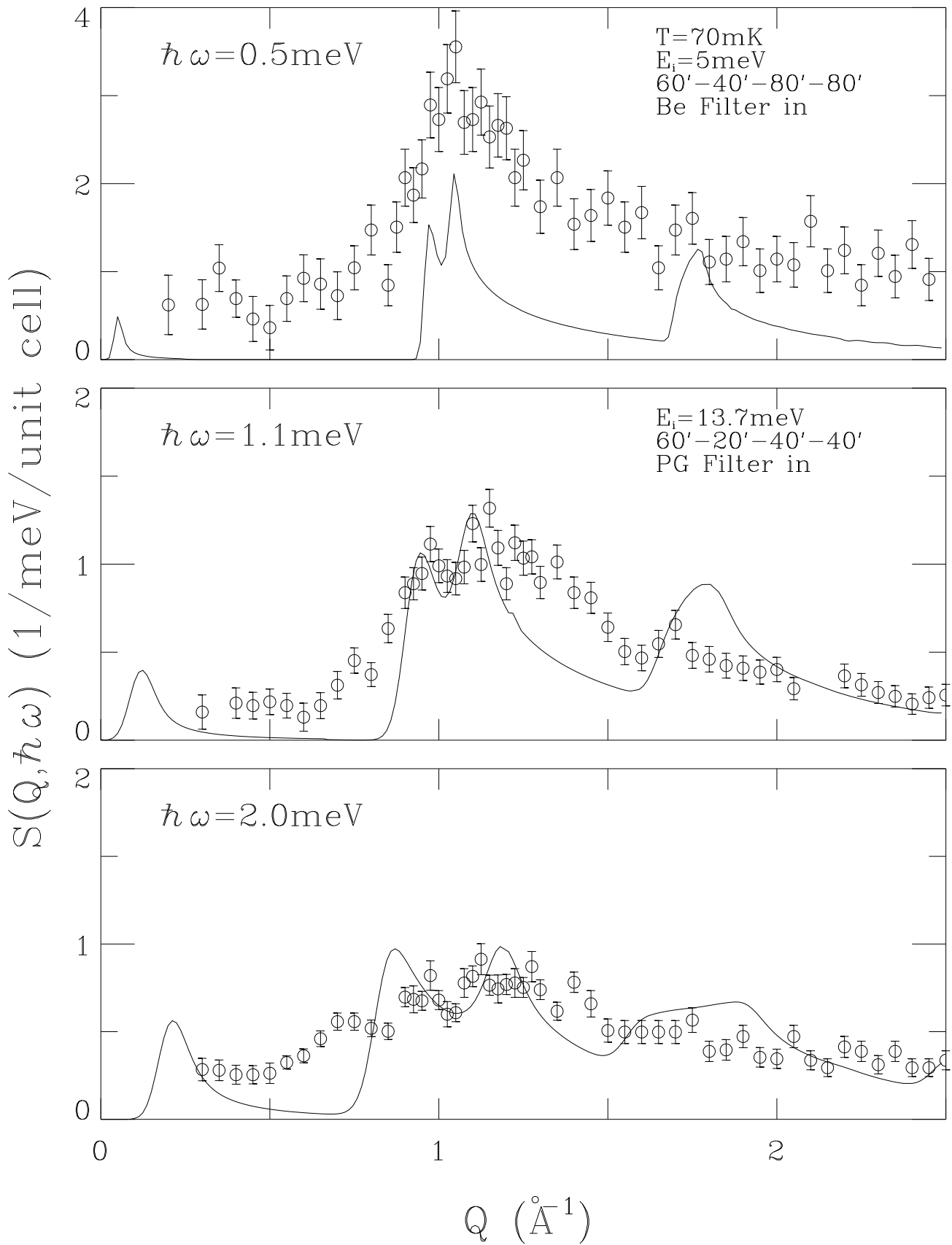
\end{document}